\documentclass[conference]{ieeeconf}

\usepackage{cite}
\usepackage{array}

\usepackage{amsthm}
\usepackage{amssymb}
\usepackage{caption}
\usepackage[gen]{eurosym}
\usepackage{subcaption}
\usepackage{siunitx}
\newcounter{thm}

\newtheorem{prob}[thm]{Problem}

\usepackage[pdftex]{graphicx}
\DeclareGraphicsExtensions{.pdf,.jpeg,.png}
\usepackage{url}

\usepackage{booktabs}
\usepackage{lettrine}

\usepackage{tikz,pgfplots}

\usetikzlibrary{shapes,arrows,positioning}
\usetikzlibrary{calc}
\usetikzlibrary{plotmarks}
\usepackage{fixmath}
\usepackage{amsmath}
\usepackage{amssymb}
\usepackage{mathrsfs}

\usepackage{algorithm}
\usepackage[noend]{algpseudocode}

\makeatletter
\def\BState{\State\hskip-\ALG@thistlm}
\makeatother

\floatname{algorithm}{Algorithm}

\usepackage{units}
\usepackage{mathtools}
\usepackage{dsfont}
\usepackage{booktabs}
\usepackage{bbm}
\usepackage{soul}
\IEEEoverridecommandlockouts

\newif\ifextendedversion 
\extendedversionfalse

\newif\ifmargincomments 
\margincommentstrue

\ifmargincomments

\else

\fi

\begin{document}
	%
	\title{
	\bf Optimal Design of Electric Micromobility Vehicles
	}
	%
	%
	%
	
	\author{Olaf Korzilius, Olaf Borsboom, Theo Hofman, Mauro Salazar
	\thanks{The authors are with the Control Systems Technology group at Eindhoven University of Technology, Eindhoven, 5600 MB, The Netherlands, {\tt\small o.a.w.korzilius@student.tue.nl, \{o.j.t.borsboom,t.hofman,m.r.u.salazar\}@tue.nl}}
	}

	\maketitle
	
	\begin{abstract}
		This paper presents a modeling and optimization framework to design battery electric micromobility vehicles, minimizing their total cost of ownership (TCO).
		Specifically, we first identify a model of the electric powertrain of an e-scooter and an e-moped consisting of a battery, a single electric motor and a transmission.
		Second, we frame an optimal joint design and control problem minimizing the TCO of the vehicles. Since this problem is nonlinear w.r.t. the motor size and the total mass of the vehicle, but convex if their value is given, we efficiently solve the problem for a range of motor sizes with an algorithm based on second-order conic programming iterating on the vehicle's mass.
		Finally, we showcase our framework on custom-created driving cycles for both vehicles in hilly and flat scenarios, providing an in-depth analysis of the results and a numerical validation with high-fidelity simulations.
		Our results show that the characteristics of the area where the vehicles are employed have a significant impact on their optimal design, whilst revealing that regenerative braking and gear-changing capabilities (as in the case of a continuously variable transmission) may not be worth implementing.

	\end{abstract}

	%
	\IEEEpeerreviewmaketitle
	
	\section{Introduction}\label{sec:introduction}
\lettrine{I}{n the} past lustrum, we have been witnessing a pervasive diffusion of electric micromobility vehicles in all major cities of the world~\cite{SimlettHolmMoeller2020}.
Whilst no formal definition exists, electric micromobility vehicles are generally regarded as low-speed, light-weight and small-size vehicles with a short driving range~\cite{YanochaAllan2019}.
In particular, this mode of transportation consists of e-bikes, e-scooters, and e-mopeds.
Such vehicles usually can be accessed either from a docking station or in designated city areas in a free-floating fashion, using scan-and-ride frameworks to rapidly unlock the vehicle and pay via a smartphone app~\cite{TuncerBrown2020}.
From a system-level perspective, these vehicles have the potential to bridge public transportation gaps in city centers, e.g., covering the first-and-last-mile legs of people's journeys in an accessible way~\cite{Clewlow2019}. What is more, compared to car-based on-demand mobility systems based on conventional vehicles, they require less parking and driving space, produce significantly less noise and harmful emission, and positively affect urban congestion~\cite{LiaoCorreia2020}.
Finally, combined with the concept of shared economies and Mobility-as-a-Service rationales, shared electric micromobility vehicles are a promising technology to explore within mobility-on-demand frameworks~\cite{Wollenstein-BetechSalazarEtAl2021}.

Given the great potential stemming from the adoption of micromobility systems, there is still plenty of room for improvement.
Besides mere regulatory issues, the average lifetime of micromobility vehicles tends to be relatively short~\cite{Granath2020} and their performance does not always meet requirements comparable with other mobility systems: For instance, in terms of gradeability performance, there are opportunities for improvement~\cite{EScooterGuideTest2021}.
Arguably, a vehicle designed for a flat city like Eindhoven may not be suited for a city with high hill-climbing requirements such as San Francisco.
Moreover, with thousands of vehicles being deployed in each city, the massive deployment scale of such transportation systems would already benefit enormously from only slight improvements in the performance of the single vehicles.
Therefore, in order to benefit the most from such a transportation mode, the design of the single vehicles should be tailored to the specific application and urban setting, calling for systematic design approaches.

\begin{figure}[!t]
	\centering
	\includegraphics[width=\columnwidth]{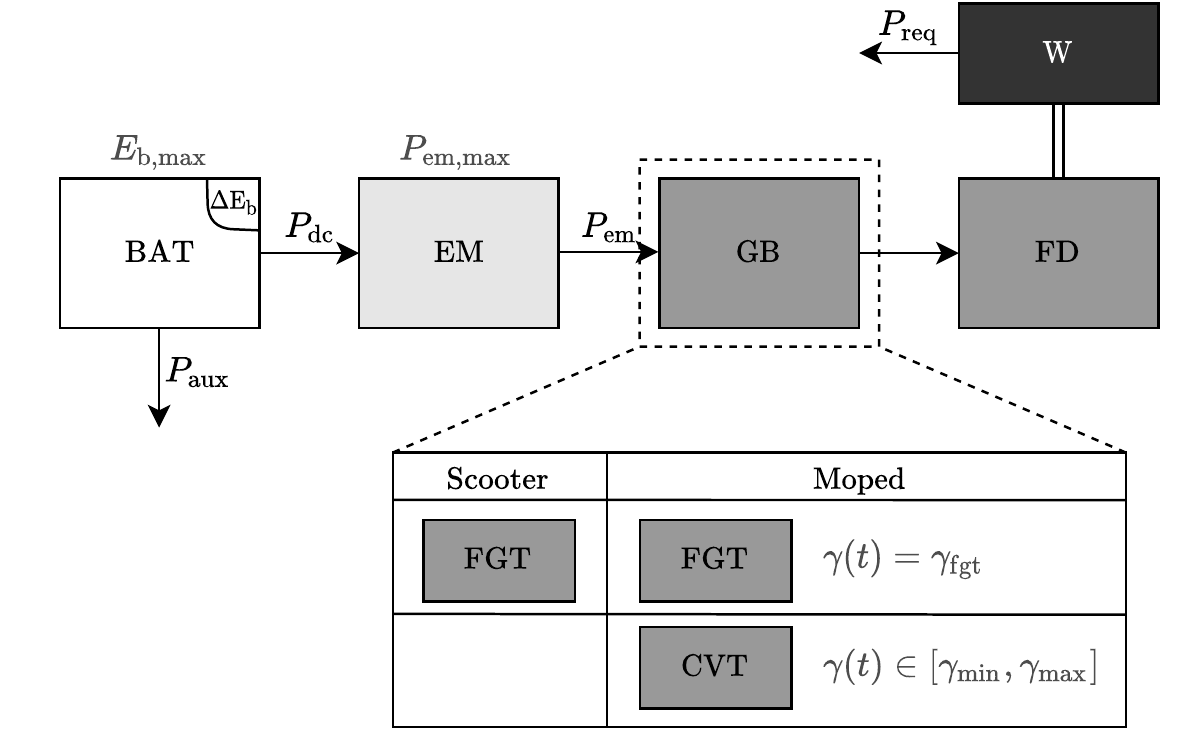}
	\caption{\small Systematic overview of the electric powertrain of the e-scooter and e-moped, consisting of a battery (BAT), an electric motor (EM), a gearbox (GB), a final drive (FD) and a driven wheel (W). The grey parameters denote the optimization variables of the components: battery size, maximum motor power, ratio of the FGT and minimum and maximum ratio of the CVT.}
	\label{fig:PT_topology}
\end{figure}

Against this background, this paper presents models and algorithms to optimize the design and operation of the powertrain of electric micromobility vehicles as shown in Fig.~\ref{fig:PT_topology}.
Specifically, this paper will focus on e-scooters with a fixed gear transmisson (FGT) and a maximum speed of \unitfrac[25]{km}{h} and e-mopeds equipped with a FGT or a continuously variable transmission (CVT) limited at \unitfrac[45]{km}{h}, whilst leaving hybrid electric human-driven vehicles such as e-bikes to future research. 

\emph{Related literature:} 
Our contribution is related to the design and control of (hybrid) electric vehicle propulsion systems and pertains to the following two research lines.
In the context of passenger cars and trucks, this design and control challenge is usually addressed with either high-fidelity nonlinear models combined with derivative-free algorithms~ \cite{EbbesenDoenitzEtAl2012, MorozovHumphriesEtAl2019, VerbruggenSilvasEtAl2020b}, or convex optimization, which sacrifices modeling accuracy in favor of time-efficient optimization algorithm~\cite{ MurgovskiJohannessonEtAl2015, PourabdollahSilvasEtAl2015, BorsboomFahdzyanaEtAl2021}.
Both these classes of methods have advantages and disadvantages. Nonetheless, they have not been applied to optimize the design and control of micromobility vehicles, which have significantly different requirements.

Narrowing the scope towards previous research on the design of e-scooters and e-mopeds, we identify the following contributions:
Whilst simulation-based approaches reveal the capabilities of (hybrid) electric two-wheelers~\cite{TsengYuEtAl2014, WalkerRoser2015,YapKarri2010}, efforts have been made to optimize the design of the electric motor of scooters and mopeds using finite-element modeling~\cite{LathaShenoySatyendraKumar2016,SongChangEtAl2010} combined with gradient-based~\cite{YangHung2013} or evolutionary algorithms~\cite{GruberBaeckEtAl2011}.
However, these methods do not jointly optimize the design and control of the full powertrain from a system-level perspective, potentially leading to sub-optimal solutions~\cite{Fathy2003}.

Summarizing, to the best of the authors' knowledge, there are no methodologies available to optimize the design and control of electric micromobility vehicles in a joint, systematic and computationally-efficient manner, whilst accounting for their application-specific requirements.

\emph{Statement of contribution:}
To bridge this gap, this paper presents an algorithm to efficiently compute the optimal design and control of an e-moped and an e-scooter in a joint fashion.
Specifically, the contribution consists of the following three items:
First, we construct an optimal control problem that aims at minimizing the total cost of ownership (TCO) of an e-moped and e-scooter in a partially-convex fashion, whilst including performance requirements on acceleration, gradeability, top speed and range. Second, we create custom driving cycles to asses the performance of the vehicles in two scenarios, namely a flat and hilly terrain. 
Third, we present an iterative, rapidly converging algorithm that solves the partially-convex problem with second-order conic programming (SOCP).

 

\emph{Organization:}
The remainder of this paper is organized as follows. In Section~\ref{sec:methodology}, we identify a partially-convex model of the powertrain and an effective solution algorithm based on SOCP to jointly optimize the design and control strategies with respect to TCO. We present numerical results on custom drive cycles with different altitude profiles in Section~\ref{sec:results}, and draw the conclusions in Section~\ref{sec:conclusion}.

	\section{Methodology}\label{sec:methodology}
This section presents a framework based on convex optimization to jointly optimize the design of the components and control strategies of the electric scooter and moped powertrain shown in Fig.~\ref{fig:PT_topology}.
First, we define the objective function of the optimization problem at hand. Subsequently, we present the model of the vehicle dynamics, gearbox, electric motor (EM), battery and vehicle mass, after which we derive the performance requirements.
Finally, we summarize the full optimization problem, devise an iterative solving algorithm and address some points of discussion regarding the proposed framework.

\subsection{Objective: Total Cost of Ownership}~\label{Sec:Objective}
The objective of the optimization problem is to minimize the TCO defined as
\par\nobreak\vspace{-5pt}
\begingroup
\allowdisplaybreaks
\begin{small}
	\begin{equation}
	\label{eq:TCO}
	J_\mathrm{TCO} = C_\mathrm{op} + C_\mathrm{comp},
	\end{equation}
\end{small}%
\endgroup
where $C_\mathrm{op}$ and $C_\mathrm{comp}$ are the operational and component costs, respectively. We define the operational cost as a function of the battery State-of-Energy (SoE) according to
\par\nobreak\vspace{-5pt}
\begingroup
\allowdisplaybreaks
\begin{small}
	\begin{align}
	\label{eq:op_cost}
	C_\mathrm{op} = \Delta E_\mathrm{b}\cdot c_\mathrm{el}\cdot \frac{D_\mathrm{max}}{D_\mathrm{cycle}},
	\end{align}
\end{small}%
\endgroup
where $c_\mathrm{el}$ is the cost of electricity, $D_\mathrm{max}$ is the distance that the vehicle can cover until battery End-of-Life, $D_\mathrm{cycle}$ is the distance covered in one driving cycle and $\Delta E_\mathrm{b}$ is the difference in battery SoE defined as
\par\nobreak\vspace{-5pt}
\begingroup
\allowdisplaybreaks
\begin{small}
	\begin{align}
	\label{eq:dEb}
	\Delta E_\mathrm{b} = E_\mathrm{b}(t_0)-E_\mathrm{b}(t_\mathrm{end}),
	\end{align}
\end{small}%
\endgroup
where $t_\mathrm{0}$ and $t_\mathrm{end}$ denote the start and end times of the driving cycle, respectively. The components' costs are
\par\nobreak\vspace{-5pt}
\begingroup
\allowdisplaybreaks
\begin{small}
	\begin{equation}
	\label{eq:componentCost}
	C_\mathrm{comp} = c_\mathrm{bat}\cdot E_\mathrm{b,max} + c_\mathrm{em}\cdot P_\mathrm{em,max} + c_\mathrm{add},
	\end{equation}
\end{small}%
\endgroup
where $ E_\mathrm{b,max}$ denotes the maximum battery capacity, $P_\mathrm{em,max}$ the maximum EM output power. Moreover, $c_\mathrm{bat}$ and $c_\mathrm{em}$ are the specific costs of the battery and electric machine (including the gearbox), respectively, and $c_\mathrm{add}$ are additional costs related to
the vehicle itself.



\subsection{Longitudinal Vehicle Dynamics}
\label{Sec:LonVehDynamics}
This study uses a quasi-static vehicle modeling approach in line with current practices~\cite{GuzzellaSciarretta2007}.
Given a driving cycle with velocity $v(t)$, acceleration $a(t)$ and gradient $\theta(t)$, the required propulsion power $P_\mathrm{req}$ is
\par\nobreak\vspace{-5pt}
\begingroup
\allowdisplaybreaks
\begin{small}
	\begin{equation}
	\label{eq:Preq}
	\begin{aligned}[b]
	& P_\mathrm{req}(t) =( m \cdot (a(t) + c_\mathrm{rr}\cdot g \cdot \mathrm{cos}(\theta(t)) + g\cdot \mathrm{sin}(\theta (t)))\\
	& + \frac{1}{2} \cdot \rho_\mathrm{a} \cdot c_\mathrm{d}  \cdot A_\mathrm{f} \cdot v(t)^2 ) \cdot v(t),
	\end{aligned}
	\end{equation}
\end{small}%
\endgroup
where $m$ is the total vehicle mass including the driver, $c_\mathrm{rr}$ is the rolling resistance coefficient, $g$ is the gravitational constant, $\rho_\mathrm{a}$ is the air density, $c_\mathrm{d}$ is the aerodynamic drag coefficient and $A_\mathrm{f}$ is the frontal area of the vehicle and driver. 
To improve readability, in the remainder of this paper we will drop time-dependence whenever clear from the context.

\subsection{Transmission}
\label{Sec:Trans}
The e-scooter and e-moped share the same powertrain topology depicted in Fig.~\ref{fig:PT_topology}, apart from the gearbox. Considering both vehicles, the gearbox transmission ratio is
\par\nobreak\vspace{-5pt}
\begingroup
\allowdisplaybreaks
\begin{small}
	\begin{equation}%
	\label{eq:gammaDef}
	\gamma(t) 
	\begin{cases}
	= \gamma_\mathrm{fgt} &\forall t \text{ if FGT}   \\
	\in [\gamma_{\mathrm{min}}, \gamma_{\mathrm{max}}]  & \forall t \text{ if CVT},  
	\end{cases}
	\end{equation}
\end{small}%
\endgroup
where $\gamma_\mathrm{fgt} > 0$ is the FGT ratio, and $\gamma_{\mathrm{min}} > 0$ and $\gamma_{\mathrm{max}} > 0$ are the lower and upper limits of the CVT ratio, respectively, which are related via a constant ratio coverage~$c_\mathrm{f}$:
\par\nobreak\vspace{-5pt}
\begingroup
\allowdisplaybreaks
\begin{small}
	\begin{equation}
	\label{eq:cvt_coverage}
	\gamma_\mathrm{max} = c_\mathrm{f}\cdot \gamma_\mathrm{min}.
	\end{equation}
\end{small}%
\endgroup

On the assumption that the efficiency of the final drive $\eta_\mathrm{fd}$ and transmission $\eta_\mathrm{gb}$ are constant, the required propulsion power can be related to the EM power $P_\mathrm{em}$:
\par\nobreak\vspace{-5pt}
\begingroup
\allowdisplaybreaks
\begin{small}
	\begin{equation}
	\label{eq:PreqPemOrigal}
	\begin{aligned}
	P_\mathrm{req} = \left\{
	\begin{array}{ll}
	\eta_\mathrm{fd} \cdot \eta_\mathrm{gb} \cdot P_\mathrm{em} & \text{~if~} P_\mathrm{em} \geq 0 \\ 
	\frac{1}{\eta_\mathrm{fd} \cdot \eta_\mathrm{gb} \cdot R_\mathrm{b}} \cdot P_\mathrm{em} -P_\mathrm{brake} & \text{~if~} P_\mathrm{em} < 0,
	\end{array}
	\right.
	\end{aligned}
	\end{equation}
\end{small}%
\endgroup
where $R_\mathrm{b}$ is the regenerative braking fraction which indicates the amount of braking power that the EM can apply to the wheels via the gearbox and final drive without destabilizing the vehicle, and $P_\mathrm{brake} \geq 0$ is the required braking power. To ensure convexity, we can follow the same reasoning as in~\cite{VerbruggenSalazarEtAl2019} and relax \eqref{eq:PreqPemOrigal} to
\par\nobreak\vspace{-5pt}
\begingroup
\allowdisplaybreaks
\begin{small}
	\begin{align}
	\label{eq:preq_relaxed}
	P_\mathrm{req} \leq \min \left( \eta_\mathrm{fd} \cdot \eta_\mathrm{gb} \cdot P_\mathrm{em},\frac{P_\mathrm{em}}{\eta_\mathrm{fd} \cdot \eta_\mathrm{gb} \cdot R_\mathrm{b}} \right).
	\end{align}
\end{small}%
\endgroup

Lastly, the rotational speed of the EM $\omega_\mathrm{em}$ is computed from the gear ratio with
\par\nobreak\vspace{-5pt}
\begingroup
\allowdisplaybreaks
\begin{small}
	\begin{align}
	\label{eq:speed}
	\omega_\mathrm{em} = \frac{v \cdot \gamma_\mathrm{fd} \cdot \gamma}{r_\mathrm{w}},
	\end{align}
\end{small}%
\endgroup
where $\gamma_\mathrm{fd}$ is the ratio of the final drive and $r_\mathrm{w}$ is the effective rolling radius of the wheel.

\subsection{Electric Motor}
\label{Sec:EM} 
\begin{figure}[!t]
	\centering
	\includegraphics[ width=1\columnwidth]{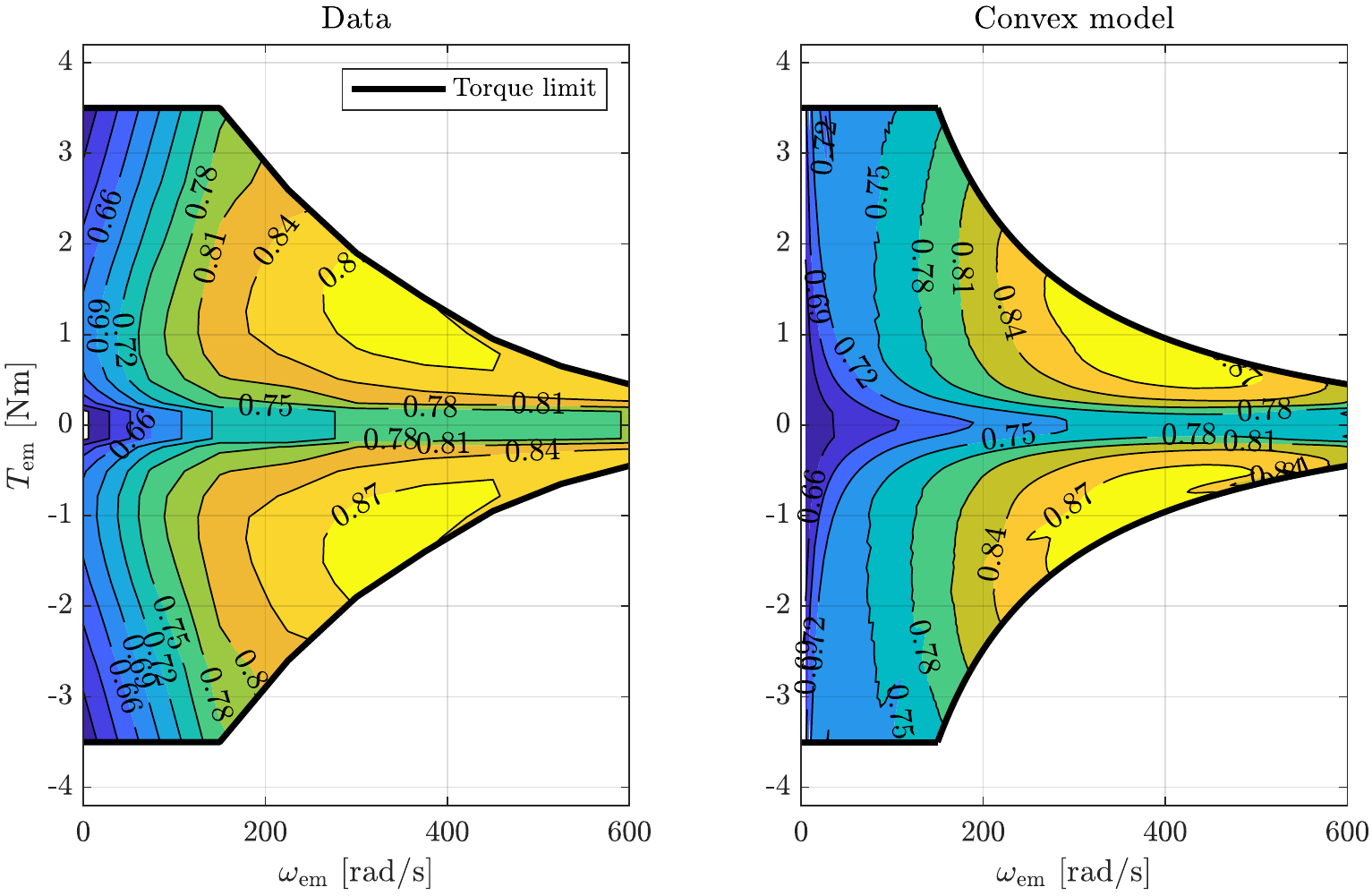}
	\caption{Efficiency map of DC-machine, including inverter efficiency (left) and convex model (right).}
	\label{fig:Efficiency map}
\end{figure}

Since the most common electric motor types among micromobility vehicles are brushed and brushless DC machines~\cite{Ulrich2005}, we identify our EM model with the DC machine data shown in Fig.~\ref{fig:Efficiency map}.


The EM input power $P_\mathrm{dc}$ is defined as 
\par\nobreak\vspace{-5pt}
\begingroup
\allowdisplaybreaks
\begin{small}
	\begin{equation}
	\label{eq:Pdc}
	P_\mathrm{dc} = P_\mathrm{em} + P_\mathrm{loss},
	\end{equation}
\end{small}%
\endgroup
where $P_\mathrm{loss}$ is the power loss.
Assuming the driving cycle and mass of the vehicle to be known in advance, the EM mechanical power $P_\mathrm{em}$ can be considered an exogenous variable.
Therefore, in line with~\cite{HurkSalazar2021}, we fit the losses as a sole function of the EM speed as the second-order polynomial
\par
\begin{figure}[!t]
	\centering
	\includegraphics[width=\columnwidth]{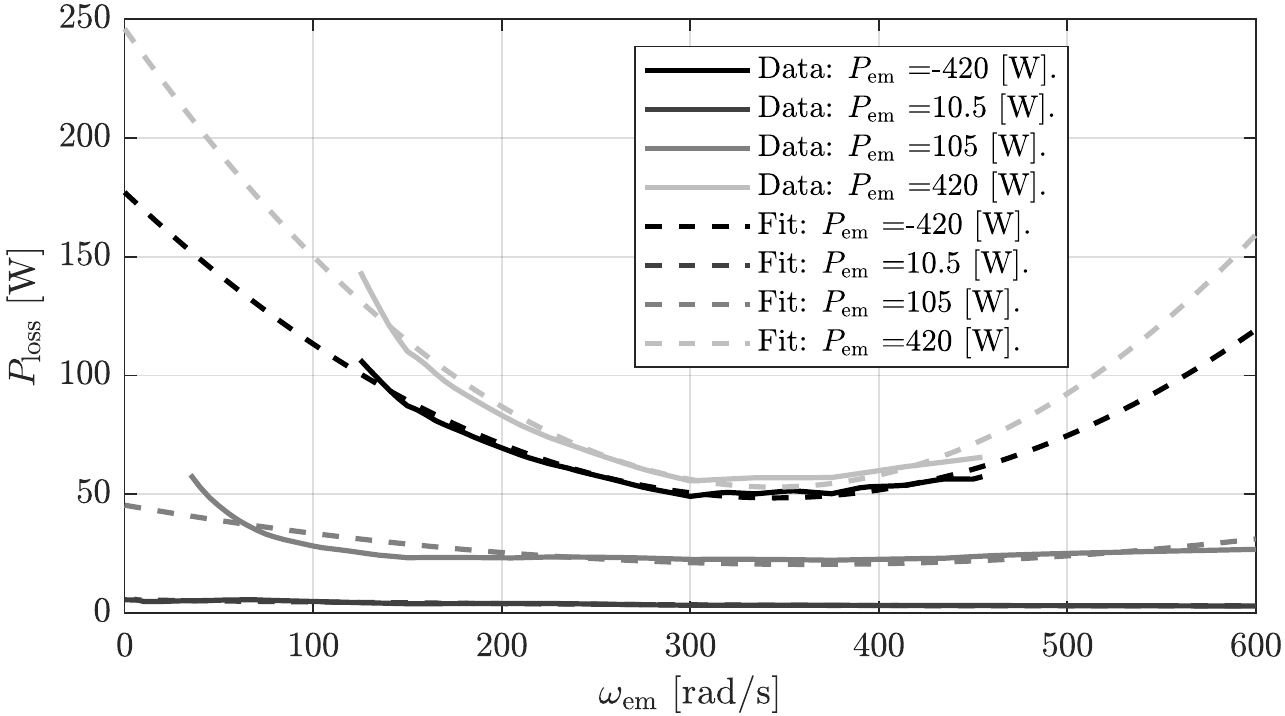}
	\caption{Quadratic fit of the power losses for four different power levels.}
	\label{fig:Powerloss_fit}
\end{figure}

\nobreak\vspace{-5pt}
\begingroup
\allowdisplaybreaks
\begin{small}
	\begin{equation}
	\label{eq:Ploss}
	P_\mathrm{loss} = a_\mathrm{1}(t) + a_\mathrm{2}(t)\cdot\omega_\mathrm{em} + a_\mathrm{3}(t)\cdot\omega_\mathrm{em}^2,
	\end{equation}
\end{small}%
\endgroup
where $a_{j}$, $j \in [1,2,3]$ are time-dependent loss coefficients that can be computed from $P_\mathrm{em}(t)$ and efficiency data. Therefore, we are able to fit the coefficients $a_{j}$ for each power loss corresponding to a range of EM power values $[-P_\mathrm{em,max},P_\mathrm{em,max}]$, creating a look-up table that returns the coefficients for a given mechanical EM power. 
The model is fitted to brushed DC machine data with a normalized root-mean-squared error (RMSE) of 1.12\%. An overview of the power loss model for four different power levels is given in Fig.~\ref{fig:Powerloss_fit}, whilst the original and modeled efficiency maps are shown in Fig.~\ref{fig:Efficiency map}.

Using the approach described above, we are able to pre-compute the EM power coefficients for each time instance if the exogenous EM power $\overline{P}_\mathrm{em}$ is given. We calculate this power from an exogenous required power request $\overline{P}_\mathrm{req}$ that is computed with \eqref{eq:Preq}, using a fixed a base vehicle mass $\overline{m}$.
Thereafter, taking into account that the EM is the only mover of the powertrain, it is possible to pre-compute the exogenous EM power $\overline{P}_\mathrm{em}$ from $\overline{P}_\mathrm{req}$ by relaxing and rewriting~\eqref{eq:PreqPemOrigal}:
\par\nobreak\vspace{-5pt}
\begingroup
\allowdisplaybreaks
\begin{small}
	\begin{equation}
	\label{eq:Pem_bar}
	\overline{P}_\mathrm{em} = \max \left(\frac{\overline{P}_\mathrm{req} }{\eta_\mathrm{gb}\cdot \eta_\mathrm{fd}}, \overline{P}_\mathrm{req}\cdot \eta_\mathrm{gb}\cdot \eta_\mathrm{fd}\cdot R_\mathrm{b}, P_\mathrm{em,min}\right),
	\end{equation}
\end{small}%
\endgroup
where $P_\mathrm{em,min} = -P_\mathrm{em,max}$. Next, the coefficients are computed and scaled as a function of maximum EM power:
\par\nobreak\vspace{-5pt}
\begingroup
\allowdisplaybreaks
\begin{small}
	\begin{align}
	a_{j}(t) =\overline{a}_{j}( \overline{P}_\mathrm{em}) \cdot \frac{P_\mathrm{em,max}}{\overline{P}_\mathrm{em,max}}~\forall j \in [1,2,3],
	\label{eq:scale_aj}
	\end{align}
\end{small}%
\endgroup
where $\overline{a}_{j}( \overline{P}_\mathrm{em})$ are the time-dependent coefficients obtained for the original EM size $\overline{P}_\mathrm{em,max}$ and from the exogenous EM power $\overline{P}_\mathrm{em}(t)$.

Similar to~\cite{VerbruggenSalazarEtAl2019}, the EM electric power is converted to a convex form by substituting~\eqref{eq:Ploss} into~\eqref{eq:Pdc} and relaxing it to
\par\nobreak\vspace{-5pt}
\begingroup
\allowdisplaybreaks
\begin{small}
	\begin{equation}
	\label{eq:Pdc_relax}
	P_\mathrm{dc} \geq P_\mathrm{em} + a_\mathrm{1}(t) + a_\mathrm{2}(t)\cdot\omega_\mathrm{em} + a_\mathrm{3}(t)\cdot\omega_\mathrm{em}^2.
	\end{equation}
\end{small}%
\endgroup
This relaxation is lossless, because our objective is to minimize the TCO, which also consists of the operational costs: Thereby, it would be inefficient to pick a larger value for $P_\mathrm{dc}$, ensuring inequality~\eqref{eq:Pdc_relax} to hold with equality~\cite{EbbesenSalazarEtAl2018}. The same reasoning applies to the other lossless relaxations that will be performed in the remainder of this paper.

The EM torque is bounded by
\par\nobreak\vspace{-5pt}
\begingroup
\allowdisplaybreaks
\begin{small}
	\begin{equation}
	\label{eq:Pem_constr_maxP}
	P_\mathrm{em} \in [-T_\mathrm{em,max}\cdot \omega_\mathrm{em}, T_\mathrm{em,max}\cdot \omega_\mathrm{em}],
	\end{equation}
\end{small}%
\endgroup
where $T_\mathrm{em,max}$ is a constant maximum torque, and the EM power is limited as
\par\nobreak\vspace{-5pt}
\begingroup
\allowdisplaybreaks
\begin{small}
	\begin{equation}
	P_\mathrm{em} \in [-k_\mathrm{m,1}\cdot \omega_\mathrm{em} - k_\mathrm{m,2}, k_\mathrm{m,1}\cdot \omega_\mathrm{em} + k_\mathrm{m,2}],
	\end{equation}
\end{small}%
\endgroup
where $k_\mathrm{m1}\leq 0$ and $k_\mathrm{m2}\geq0$ are the maximum power coefficients subject to identification~\cite{BorsboomFahdzyanaEtAl2021}. Similar to \eqref{eq:scale_aj}, we scale the maximum torque and coefficients as a function of maximum EM power using 
\par\nobreak\vspace{-5pt}
\begingroup
\allowdisplaybreaks
\begin{small}
	\begin{align}
	T_\mathrm{em,max}  = \overline{T}_\mathrm{em,max} \cdot \frac{P_\mathrm{em,max}}{\overline{P}_\mathrm{em,max}}\\
	k_\mathrm{m,i} = \overline{k}_\mathrm{m,i} \cdot \frac{P_\mathrm{em,max}}{\overline{P}_\mathrm{em,max}}~\forall i \in [1,2].
	\label{eq:kj}
	\end{align}
\end{small}%
\endgroup

Lastly, using \eqref{eq:speed} and given that $\omega_\mathrm{em} \leq \omega_\mathrm{em,max}$, the maximum EM speed constraint can be rewritten as a function of $\gamma$, which results in

\par\nobreak\vspace{-5pt}
\begingroup
\allowdisplaybreaks
\begin{small}
	\begin{equation}
	\label{eq:Overspeeding_constr}
	\gamma(t) \leq \frac{\omega_\mathrm{em,max}\cdot r_\mathrm{w}}{ \gamma_\mathrm{fd}\cdot v}.
	\end{equation}
\end{small}%
\endgroup

\subsection{Battery Pack}
\label{Sec:BAT}
We model the battery using the same approach that is discussed in~\cite{VerbruggenSalazarEtAl2019}, where the battery is modeled according to a standard equivalent circuit with internal resistance $R$ and open-circuit voltage as a function of battery SoE $V_\mathrm{oc}(E_\mathrm{b})$. We compute the power at the battery terminals with
\par\nobreak\vspace{-5pt}
\begingroup
\allowdisplaybreaks
\begin{small}
	\begin{equation}
	\label{eq:Pbat}
	P_\mathrm{b} = P_\mathrm{dc} + P_\mathrm{aux},
	\end{equation}
\end{small}%
\endgroup
where $P_\mathrm{aux}$ is a constant auxiliary power. Furthermore, the internal battery power $P_\mathrm{i}$, which is responsible for the change in battery SoE, is related to $P_\mathrm{b}$ with

\par\nobreak\vspace{-5pt}
\begingroup
\allowdisplaybreaks
\begin{small}
	\begin{equation}
	\label{eq:pi_rewritten}
	(P_\mathrm{i} - P_\mathrm{b}) \cdot P_\mathrm{oc} =  P_\mathrm{i}^2.
	\end{equation}
\end{small}%
\endgroup
where $P_\mathrm{oc}$ is the open circuit power which we write as a function of $E_\mathrm{b}$ and $E_\mathrm{b,max}$ as
\par\nobreak\vspace{-5pt}
\begingroup
\allowdisplaybreaks
\begin{small}
	\begin{equation}
	\label{eq:Poc_fit}
	P_\mathrm{oc} = a_{1}\cdot E_\mathrm{b} + a_{2}\cdot E_\mathrm{b,max}.
	\end{equation}
\end{small}%
\endgroup
where $a_\mathrm{1}$ and $a_\mathrm{2}$ are coefficients subject to identification.
Then, in order to achieve convexity, we relax \eqref{eq:pi_rewritten} and write it as a second-order order conic constraint:
\par\nobreak\vspace{-5pt}
\begingroup
\allowdisplaybreaks
\begin{small}
	\begin{equation}
	\begin{Vmatrix}
	2\cdot P_\mathrm{i}\\
	P_\mathrm{i} - P_\mathrm{b} - P_\mathrm{oc}
	\end{Vmatrix}_\mathrm{2}    \leq P_\mathrm{i} - P_\mathrm{b} + P_\mathrm{oc}.
	\end{equation}
\end{small}%
\endgroup
We limit the internal battery power as
\par\nobreak\vspace{-5pt}
\begingroup
\allowdisplaybreaks
\begin{small}
	\begin{equation}
	\label{eq:Pimax}
	P_\mathrm{i} \in  [-P_\mathrm{i,max},P_\mathrm{i,max} ],
	\end{equation}
\end{small}%
\endgroup
where $P_\mathrm{i,max}$ is the maximum internal power that is related to $E_\mathrm{b}$ with the affine function
\par\nobreak\vspace{-5pt}
\begingroup
\allowdisplaybreaks
\begin{small}
	\begin{align}
	P_\mathrm{i,max} = b_\mathrm{1}\cdot E_\mathrm{b} + b_\mathrm{2}\cdot E_\mathrm{b,max},
	\end{align}
\end{small}%
\endgroup
where the coefficients $b_\mathrm{1}$ and $b_\mathrm{2}$ are again subject to identification. In addition, we limit the battery SoE with
\par\nobreak\vspace{-5pt}
\begingroup
\allowdisplaybreaks
\begin{small}
	\begin{equation}
	E_\mathrm{b} \in [\zeta_\mathrm{b,min}, \zeta_\mathrm{b,max} ] \cdot E_\mathrm{b,max},
	\end{equation}
\end{small}%
\endgroup
where $\zeta_\mathrm{b,min}$ and $\zeta_\mathrm{b,max}$ are the minimum and maximum SoE levels, respectively. Lastly, the internal battery dynamics are given by
\par\nobreak\vspace{-5pt}
\begingroup
\allowdisplaybreaks
\begin{small}
	\begin{equation}
	\label{eq:Eb}
	\frac{\mathrm{d}}{\mathrm{d}t} E_\mathrm{b} = -P_\mathrm{i}.
	\end{equation}
\end{small}%
\endgroup

The battery model is fitted to the data from a pack of INR18650-25R lithium ion cells in 13-series-1-parallel formation with a normalized RMSE of 0.53\%~\cite{KimLeeEtAl2019}.
This configuration gives the pack a nominal voltage of \unit[48]{V} and capacity of \unit[2.5]{Ah}.

\subsection{Mass}
\label{Sec:mass}
The gross vehicle mass is computed using
\par\nobreak\vspace{-5pt}
\begingroup
\allowdisplaybreaks
\begin{small}
	\begin{equation}
	\label{eq:mv}
	m = m_\mathrm{d} + m_\mathrm{v}
	\end{equation}
\end{small}%
\endgroup
where $m_\mathrm{d}$ is the driver weight and $m_\mathrm{v}$ is the vehicle mass, which is calculated using
\par\nobreak\vspace{-5pt}
\begingroup
\allowdisplaybreaks
\begin{small}
	\begin{equation}
	\label{eq:mv2}
	m_\mathrm{v} = m_\mathrm{em} + m_\mathrm{bat} + m_\mathrm{gb} + m_\mathrm{f},
	\end{equation}
\end{small}%
\endgroup
in which $m_\mathrm{em}$ is the electric EM mass, $m_\mathrm{gb}$ is the transmission mass, $m_\mathrm{bat}$ is the battery mass and $m_\mathrm{f}$ is the frame mass. In contrast to conventional vehicles, the driver mass has to be included for the optimization of micromobility vehicles, because it often is greater than or equal to the vehicle weight.
We adjust the mass of the three scalable components of the powertrain as follows:
First, the EM mass is scaled linearly with the maximum EM power using 
\par\nobreak\vspace{-5pt}
\begingroup
\allowdisplaybreaks
\begin{small}
	\begin{equation}
	m_\mathrm{em} = \rho_\mathrm{em} \cdot P_\mathrm{em,max},
	\end{equation}
\end{small}%
\endgroup
where $\rho_\mathrm{em}$ is the specific EM mass per power unit, which also incorporates the mass accounted by the power electronics.
Secondly, the battery mass is scaled with the maximum battery capacity using
\par\nobreak\vspace{-5pt}
\begingroup
\allowdisplaybreaks
\begin{small}
	\begin{equation}
	\label{eq:mbat}
	m_\mathrm{bat} = \rho_\mathrm{bat} \cdot E_\mathrm{b,max},
	\end{equation}
\end{small}%
\endgroup
where $\rho_\mathrm{bat}$ is the specific battery weight per unit of energy.
Finally, in line with~\cite{VerbruggenSalazarEtAl2019}, we scale the transmission mass quadratically with respect to the gear ratio and relax it as
\par\nobreak\vspace{-5pt}
\begingroup
\allowdisplaybreaks
\begin{small}
	\begin{align}
	\label{eq:trasmission_mass}
	m_\mathrm{gb} \geq \left\{
	\begin{array}{ll}
	\rho_\mathrm{fgt} \cdot \gamma_\mathrm{fgt}^2 & \text{if FGT} \\
	m_\mathrm{cvt,base} + \rho_\mathrm{cvt} \cdot \gamma_\mathrm{max}^2 & \text{if CVT},\\
	\end{array}
	\right.
	\end{align}
\end{small}%
\endgroup
where $\rho_\mathrm{fgt}$ and $\rho_\mathrm{cvt}$ are the specific gearbox weights of the FGT and CVT, respectively, and $m_\mathrm{cvt,base}$ is the base mass of the CVT.

\subsection{Performance Requirements}
\label{Sec:PerfReq} 
As mentioned in Section~\ref{sec:introduction}, we include performance constraints in the optimization problem to ensure that the gradeability, top speed, acceleration and range of the vehicles are acceptable. We define the gradeability requirement for the vehicles as
\par\nobreak\vspace{-5pt}
\begingroup
\allowdisplaybreaks
\begin{small}
	\begin{align}
			\label{eq:grad_constr_e_scooter}
			m \cdot g \cdot \mathrm{sin}(\theta_\mathrm{start}) \cdot r_\mathrm{w} \leq 
			 \eta_\mathrm{gb} \cdot \eta_\mathrm{fd} \cdot T_\mathrm{em,max}\cdot \gamma_\mathrm{fd} \cdot \gamma_\mathrm{x}
	\end{align}
\end{small}%
\endgroup
where $\gamma_\mathrm{x} = \gamma_\mathrm{fgt}$ for the FGT, $\gamma_\mathrm{x} = \gamma_\mathrm{max}$ for the CVT and $\theta_\mathrm{start}$ is the gradient from which the vehicle should be able to start driving from standstill. 
Subsequently, we ensure that the vehicle is able to drive at top speed on a flat road, without passing the torque limit, using
\begingroup
\allowdisplaybreaks
\begin{small}
	\begin{equation}
	\label{eq:vel_req}
	\begin{aligned}[b]
	&  T_\mathrm{req}(v_\mathrm{max}) \leq \min(T_\mathrm{em,max}\cdot \eta_\mathrm{fd}\cdot \eta_\mathrm{gb}\cdot \gamma_\mathrm{x}\cdot \gamma_\mathrm{fd}, \\
	& (k_\mathrm{m,1}\cdot\gamma_\mathrm{x}\cdot \gamma_\mathrm{fd}~+k_\mathrm{m,2}\cdot \frac{r_\mathrm{w}}{v_\mathrm{max}})\cdot \eta_\mathrm{fd}\cdot\eta_\mathrm{gb}),
	\end{aligned}
	\end{equation}
\end{small}%
\endgroup
where $\gamma_\mathrm{x}  =\gamma_\mathrm{fgt}$ for the FGT, $\gamma_\mathrm{x}  =\gamma_\mathrm{min}$ for the CVT, $v_\mathrm{max}$ is the top speed and  $T_\mathrm{req}$ is the required torque to drive at the top speed, which is calculated as
\par\nobreak\vspace{-5pt}
\begingroup
\allowdisplaybreaks
\begin{small}
	\begin{equation}
	T_\mathrm{req}(v_\mathrm{max}) = \frac{P_\mathrm{req}(v_\mathrm{max})}{v_\mathrm{max}} \cdot r_\mathrm{w}.
	\end{equation}
\end{small}%
\endgroup
Moreover, we ensure that the EM power is high enough such that the vehicles are able to reach their top speed within an acceleration time $t_\mathrm{acc}$ with

\begingroup
\allowdisplaybreaks
\begin{small}
	\begin{equation}
	\label{eq:acc_constraint}
	P_\mathrm{em,max}\cdot \eta_\mathrm{gb}\cdot \eta_\mathrm{fd} \geq \frac{v_\mathrm{max}^2\cdot m}{t_\mathrm{acc}}.
	\end{equation}
\end{small}%
\endgroup
Lastly, the battery size is determined from the expected range, energy consumption during one cycle, and the minimum charge level  using  


\par\nobreak\vspace{-5pt}
\begingroup
\allowdisplaybreaks
\begin{small}
	\begin{equation}
	\label{eq:Bat_req}
	E_\mathrm{b,max} \geq \frac{\Delta E_\mathrm{b}}{1-\zeta_\mathrm{b,min}}\cdot \frac{D_\mathrm{exp}}{D_\mathrm{cycle}},
	\end{equation}
\end{small}%
\endgroup
where $D_\mathrm{exp}$ is the expected range.

\subsection{TCO Based Component Sizing and Control Problem}\label{sec:optprob}



In this section, we present the optimization problem, summarizing the objective function and modeling constraints derived in the previous sections. Using second-order conic programming (SOCP), we solve the joint design and control problem for the state variable $x = E_\mathrm{b}$, the control variables $u = \{P_\mathrm{em},\gamma(t)\}$, and the design variables \mbox{$p = \{P_\mathrm{em,max}, E_\mathrm{b,max}, \gamma_i\}$},  where $i=\mathrm{fgt}$ and $i = \{\mathrm{min},\mathrm{max}\}$ for the FGT and CVT, respectively.
\begin{prob}[]\label{prob:CVX}
	The minimum TCO, component sizes and control strategies are the solution of
	\par\nobreak\vspace{-5pt}
	\begingroup
	\allowdisplaybreaks
	\begin{small}
		\begin{equation*}
		\begin{aligned}
		\min\, & J_\mathrm{TCO}\\
		\text{s.t. }&\eqref{eq:TCO}-\eqref{eq:cvt_coverage},~ \eqref{eq:preq_relaxed},~\eqref{eq:speed},\\ &\eqref{eq:Pem_bar}-\eqref{eq:Pbat},~ \eqref{eq:Poc_fit}-\eqref{eq:Bat_req}.
		\end{aligned} 
		\end{equation*}
	\end{small}%
	\endgroup
\end{prob}
Problem~\ref{prob:CVX} is not completely convex because (i) the EM scaling results in a bi-linearity and (ii) the optimized mass is not equal to the mass that is used to pre-compute the EM coefficients, as it is dependent on the components' size.
To this end, we present an iterative solution algorithm that circumvents these issues and consists of two main aspects.
First, in order to ensure convexity, we fix the EM size and the mass of the powertrain, including the battery and the transmission, allowing us to pre-compute the EM coefficients.
Second, we vary the EM size for a vector of maximum EM powers with a fine discretization and solve Problem~\ref{prob:CVX} for each of these given values by iterating on the mass of the powertrain, which would result from the size of the battery and the transmission.

To go into more detail of this solving procedure we refer to Algorithm~1, in which we search the optimal vehicle mass $m_\mathrm{v}^*$ for each given value of $P_\mathrm{em,max}$.
In particular, we take an initial guess $m_\mathrm{v,0}$ for the vehicle mass, we pre-compute the EM coefficients $a_j$ and solve Problem~\ref{prob:CVX} as an SOCP, which returns an optimal battery and transmission sizing.
Using this solution, we can update the vehicle mass and follow the same procedure.
We stop the iterations when the mass of the current iteration coincides with the mass of the previous iteration, up to the tolerance $\epsilon$.
\begin{table}
	\centering
	\label{alg:Algorithm 1}
	\begin{tabular}{l}\toprule
		\textbf{Algorithm 1}: iterative solving procedure\\ \hline 
		Given $P_\mathrm{em,max}$:  \vspace{1mm}\\ 	
		Initial guess: $m_\mathrm{v}^* = m_\mathrm{v,0}, \overline{m}_\mathrm{v} = 0$\\ 
		\textbf{while} $||\overline{m}_\mathrm{v}-m_\mathrm{v}^*|| \geq \epsilon$ \\
		$~~\overline{m}_\mathrm{v} = m_\mathrm{v}^*$\\
		$~~\overline{m} = \overline{m}_\mathrm{v} + m_\mathrm{d}$\\ 
		$~~\mathrm{Precompute} ~ a_\mathrm{j}(t)$\\
		$~~m_\mathrm{v}^* \xleftarrow[]{} \mathrm{solve\; SOCP}$ \\
		\textbf{end}\\
		$m^*=m_\mathrm{v}^*+m_\mathrm{d}$\\\hline
	\end{tabular}
\end{table}

\subsection{Discussion}
\label{Sec:Discussion}
A few comments are in order.
First, we assume the EM losses and maximum torque to scale linearly with the maximum EM power $P_\mathrm{em,max}$, such that the efficiency map values remain unchanged, which is common practice in such sizing studies. Since the EM torque scales linearly in EM length, the EM mass is also scaled linear with respect to EM power~\cite{Thiringer2017}. However, as we are separately solving for different EM sizes, our framework allows for more accurate sizing models.
Moreover, we scale the battery by adding branches in parallel, which ensures the open-circuit voltage to remain constant. Thereby, the battery pack was fitted with respect to a pack with an output voltage of \unit[48]{V}, which corresponds to micromobility applications found in industry~\cite{ScooterBatteryVoltage2021}.
Lastly, the solution found by Algorithm~1 does not have global optimality guarantees, because Algorithm 1 is solving Problem~\ref{prob:CVX}, which, for a fixed $P_\mathrm{em,max}$, is still nonlinear with respect to $m_\mathrm{v}$.
 However, the numerical convergence analysis provided
\ifextendedversion
in Fig.~\ref{fig:convergence_analysis_sooter} in the Appendix shows promising results.
\else
in the extended version of this paper shows promising results~\cite{KorziliusBorsboomEtAl2021}. 
\fi
 In addition, as shown in Section~\ref{sec:results} below, we also solved Problem~\ref{prob:CVX} with nonlinear programming methods. While we obtained the same results for the FGT-equipped vehicles, the CVT-case did not converge, validating and motivating the proposed approach.



	\section{Numerical Results}\label{sec:results}
\begin{figure}[!t]
	\centering
	\includegraphics[ width=1\columnwidth]{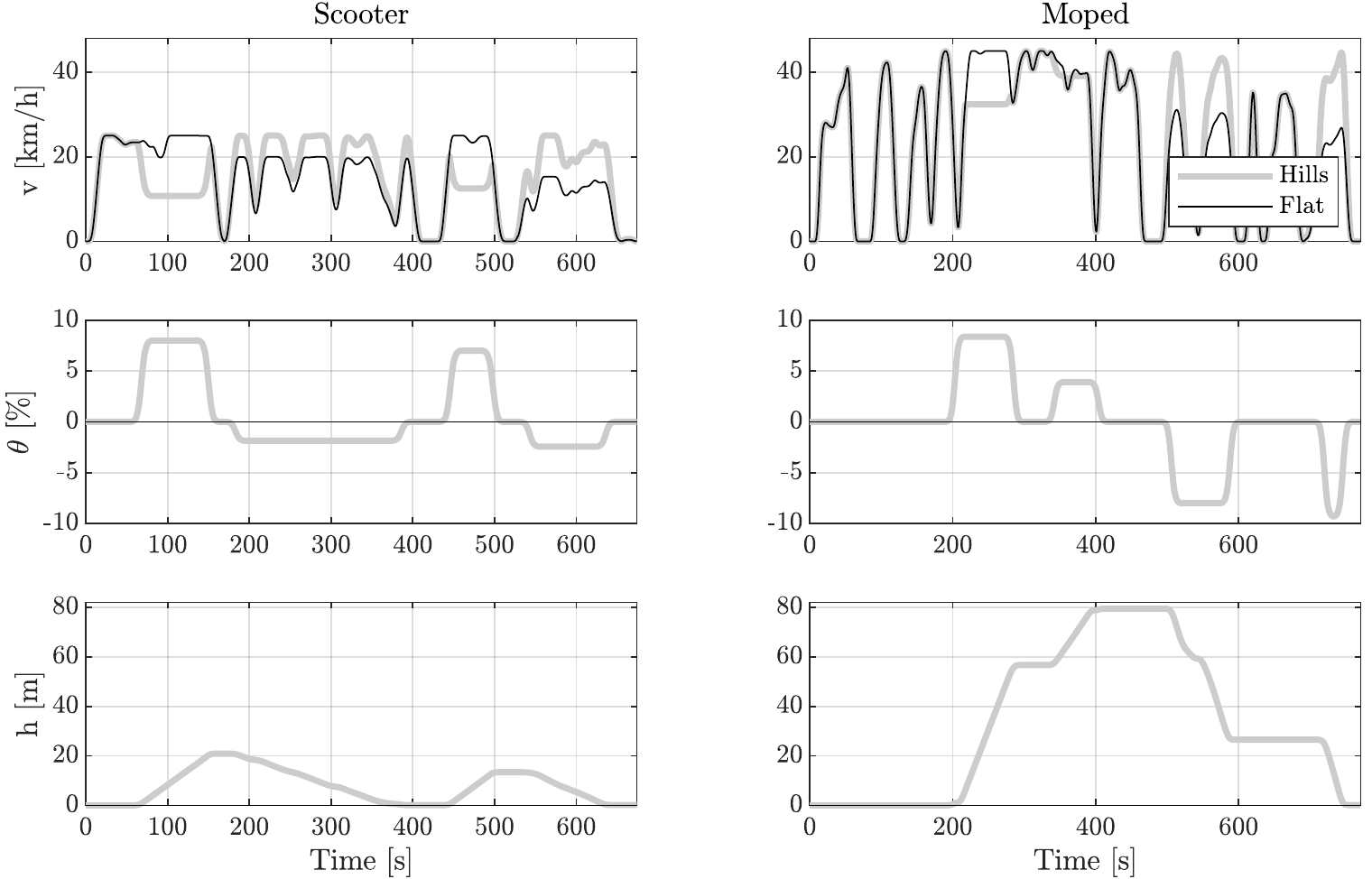}
	\caption{The driving cycles used for the micromobility vehicles. }
	\label{fig:DrivingCycles}
\end{figure}

This section presents the numerical results obtained when leveraging our framework proposed in Section~\ref{sec:methodology} above to optimize the design and control of an e-scooter and an e-moped.
In line with common practices for the design and control optimization of battery and hybrid-electric vehicles~\cite{GuzzellaSciarretta2007}, we use driving cycles consisting of exogenous velocity and gradient profiles.
Official test cycles for micromobility vehicles are not yet available. Therefore, for the e-scooter, we measure the driving cycle by completing a representative driving mission in Eindhoven and using the log data on speed, distance and height registered with a GPS-based application like Strava~\cite{Strava2021}.
For the e-moped, we use the transient phase of an urban cycle (EPA Urban Dynamometer Driving Schedule~\cite{EPA2021}) which is speed-limited to \unitfrac[45]{km}{h}.
As both driving cycles result from a flat terrain, we create hilly scenarios by including a synthetic gradient profile with equal start and end altitudes and adjust the speed profile to make it realistic.
The resulting cycles are shown in Fig.~\ref{fig:DrivingCycles}.

In order to solve Problem~\ref{prob:CVX}, we discretize it with the Euler Forward method and a sampling time of \unit[1]{s}.
As mentioned in Section~\ref{sec:optprob}, Algorithm~1 iteratively solves Problem~\ref{prob:CVX} with respect to the vehicle mass, using SOCP for fixed values of $P_\mathrm{em,max} \in$ [300, 800] \unit{W} for the scooter and $P_\mathrm{em,max} \in$ [2000, 3000] \unit{W} for the moped, with a tolerance $\epsilon$ of \unit[0.001]{kg}.
Thereby, we parse the problem with YALMIP~\cite{Loefberg2004} and solve it using MOSEK~\cite{MosekAPS2010}.
As the TCO-contribution of the components' costs is significantly higher than the cost of electricity (as shown in Table \ref{Tab:Simulation results scooter} and \ref{Tab:Simulation results_moped}), the optimal motor size is generally located close to the smallest feasible motor power. We observed an average computation time of \unit[20]{s} per motor size (considering that Algorithm 1 typically converges to an optimal mass in three iterations) on a machine with an Intel\textregistered~Core\texttrademark~i7-4710MQ CPU and \unit[8]{GB} of RAM. The total computation time, including the optimal EM sizing, equals about \unit[820]{s} for the e-scooter and e-moped.

Beside leveraging our presented solution algorithm, we also solve Problem~\ref{prob:CVX} directly as a nonlinear program (NLP). Thereby, whilst for the FGT-equipped vehicles we obtain similar computation times and identical results, convergence cannot be achieved for the CVT-equipped e-moped, motivating and validating the use of the proposed Algorithm~1 to solve Problem~\ref{prob:CVX}.




\subsection{Case Study: Flatland Compared to Hills}
To analyze the impact of the area of employment on the optimal vehicle design, we use the driving cycles shown in Fig.~\ref{fig:DrivingCycles}. Thereby, for each vehicle-type we consider a flat one (e.g., representing Eindhoven) and a hilly one (e.g., representing San Francisco).
The results of the case study are presented in Table~\ref{Tab:Simulation results scooter} and Table~\ref{Tab:Simulation results_moped}, and are obtained using the simulation parameters provided
\ifextendedversion
in the Appendix.
\else
in the extended version of this paper~\cite{KorziliusBorsboomEtAl2021}.
\fi
\begin{small}
	\begin{table}[!t]
		\centering
		\caption{Simulation results for the e-scooter.}
		\label{Tab:Simulation results scooter}\scriptsize
		\begin{tabular}{l| l l}\hline
			& \textit{Flat} & \textit{Hills}    \\ \hline
			$J(\mathrm{TCO})$ [\si{\euro}]     &  296   & 318   \\ 
			$C_\mathrm{comp}$ [\si{\euro}]     &  272   & 293    \\ 
			$C_\mathrm{el}$ [\si{\euro}]     &   24 & 25   \\ 
			$P_\mathrm{m,max}$ [\si{W}]          &  590  &  640 \\ 
			$E_\mathrm{b,max}$ [\si{Wh}]         &  435  & 491   \\ 
			$m_\mathrm{v}$ [\si{kg}]         &  12.7 &  13.1  \\ 
			$\gamma$ [-]          &  5.91 &    6.77   \\ \hline
		\end{tabular}
	\end{table}
\end{small}
\begin{small}
	\begin{table}[!t]
		\centering
		\caption{Simulation results of the e-moped for both scenarios and transmission technologies.}\scriptsize
		\label{Tab:Simulation results_moped}
		\begin{tabular}{l| l l l}\hline
			\textbf{Scenario}	  &   \textbf{Transmission} &  &    \\ \hline
			\textit{Flat}	& \textit{FGT} 	& \textit{CVT}    & \\ 
			$J(\mathrm{TCO})$ [\si{\euro}]          &  1713  & 2018 & (+17.8\%)\\ 
			$C_\mathrm{comp}$ [\si{\euro}]          &  1175& 1411  & (+16.7\%)\\ 
			$C_\mathrm{el}$ [\si{\euro}]          &  538  & 607  & (+13.9\%)\\
			$P_\mathrm{m,max}$ [\si{W}]          &  2370  & 2550 & (+7.59\%)\\ 
			$E_\mathrm{b,max}$ [\si{Wh}]         &  2549  & 2874 &(+12.8\%)\\  
			$m_\mathrm{v}$ [\si{kg}]         &  75.1  & 78.2  & (+4.13\%)\\
			$\gamma_\mathrm{fgt}$ [-]          &  5.03  & -   & \\
			$\gamma_\mathrm{min}$ [-]          &  - & 2.80    & \\ 
			$\gamma_\mathrm{max}$ [-]          &  -  & 7.57   & \\  \hline
			\textit{Hills}      &    &  &  \\ 
			$J(\mathrm{TCO})$ [\si{\euro}]      &  1857  & 2197  &  (+18.3\%)\\ 
			$C_\mathrm{comp}$ [\si{\euro}]          &  1263  & 1516 & (+20.0\%)\\ 
			$C_\mathrm{el}$ [\si{\euro}]          &  594  & 681 & (+14.7\%)\\
			$P_\mathrm{m,max}$ [\si{W}]          &  2490 & 2580 & (+3.6\%)\\ 
			$E_\mathrm{b,max}$ [\si{Wh}]         &  2815  & 3227 &  (+14.6\%)\\ 
			$m_\mathrm{v}$ [\si{kg}]         &  76.9  & 80.0   &  (+4.03\%)\\ 
			$\gamma_\mathrm{fgt}$ [-]          & 5.56  &  -    & \\ 
			$\gamma_\mathrm{min}$ [-]          &  -  & 2.86    & \\ 
			$\gamma_\mathrm{max}$ [-]          &  -  &7.73    & \\  \hline
		\end{tabular}
	\end{table}
\end{small}

Table~\ref{Tab:Simulation results scooter} shows that in the hilly scenario, the optimal TCO of the e-scooter increases by 7.4\% with respect to the flat scenario. 
In particular, the components' and electrical costs rise by 7.7\% and 4.2\%, respectively.
Similar trends are observed for the e-moped, whereby the costs increase significantly.
Specifically, the component and electrical costs for the FGT-equipped vehicle increase by 6.9\% and 10.4\%, respectively, whereas they increase by 7.4\% and 12.2\% for the CVT-equipped one.
Overall, we observe that the type of terrain has a significant influence on the optimal component sizes and should be considered prior to deployment.
Moreover, we observe that the influence of the components' cost on the TCO is about a factor of ten larger than the electricity costs for the e-scooter, whilst this difference is reduced to a factor of two for the e-moped.

In addition, the fixed-gear ratio of the e-scooter increases from 5.91 for the flat case to 6.77 in the hilly terrain.
This difference is caused by the larger EM size, which allows the operating points to be shifted towards a more efficient region at higher speeds, without exceeding the EM torque limits. The increase in gear ratio for the FGT-moped in the hills can be explained following a similar reasoning.
In contrast, the maximum and minimum allowable CVT ratios do not change substantially, because they are related to performance requirements, which are only affected by the limited increase in vehicle mass.
A further analysis of the difference between both transmission technologies can be found in
\ifextendedversion
the Appendix.
\else
the extended version of this paper~\cite{KorziliusBorsboomEtAl2021}.
\fi

Furthermore, Table~\ref{Tab:Simulation results_moped} shows that the TCO increases by more than 17\% for the CVT powertrain with respect to the FGT powertrain in both scenarios. 
Although the EM operating points of the CVT are placed in more efficient regions, the resulting decrease in energy consumption does not compensate for the lower transmission efficiency and higher mass and implementation costs.
Moreover, the lower CVT-efficiency requires the motor to be 3.6\% larger in order to meet the constraints, which also increases the TCO. All in all, these factors make the FGT-equipped e-moped the more cost-effective solution. 

\begin{figure}[!t]
	\centering
	\includegraphics[ width=1\columnwidth]{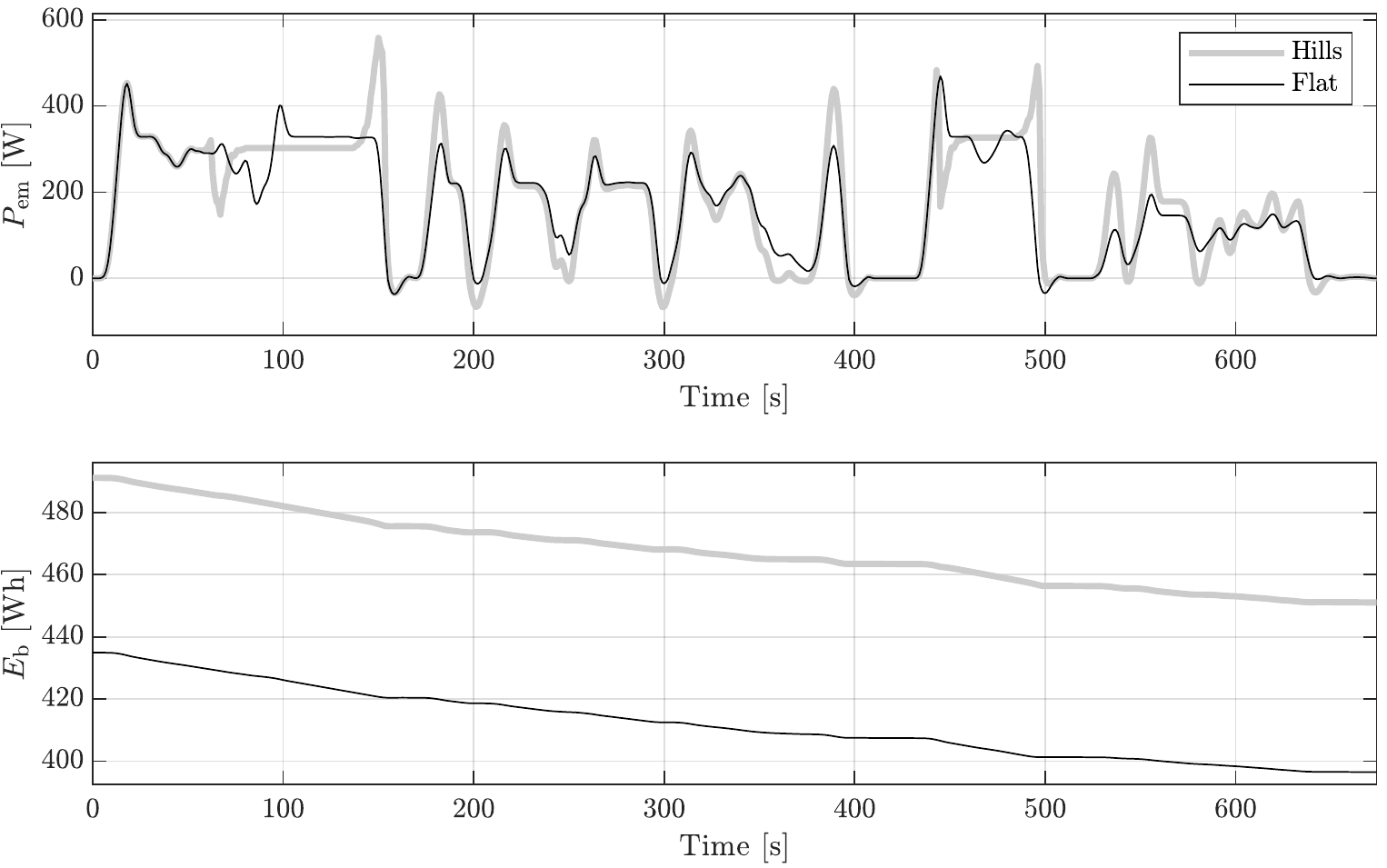}
	\caption{Mechanical EM power and battery SoE of the e-scooter in both scenarios.}
	\label{fig:controls_scooter}
\end{figure}

\begin{figure}[!t]
	\centering
	\includegraphics[ width=1\columnwidth]{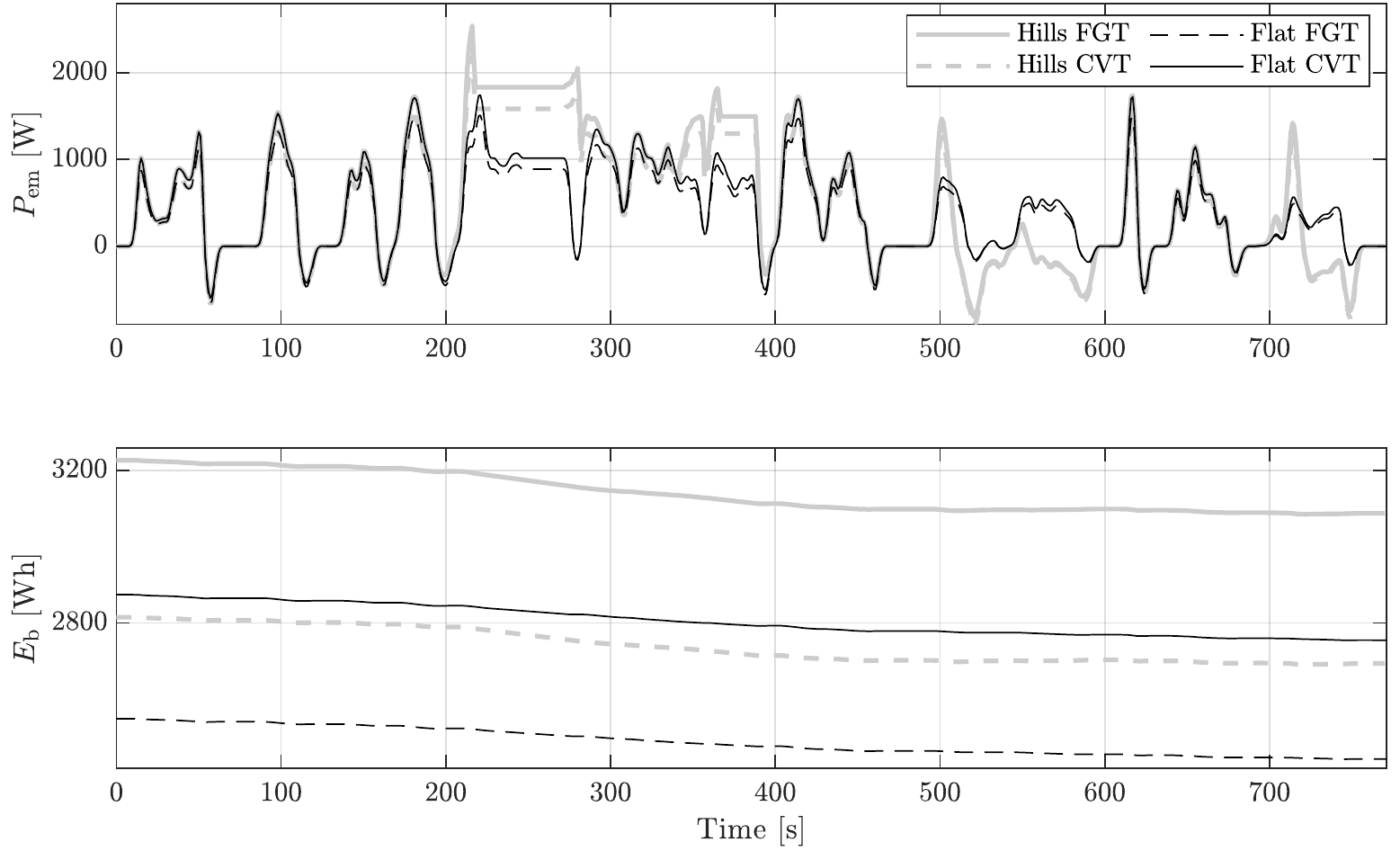}
	\caption{Mechanical EM power and battery SoE of the e-moped for both scenarios and transmission technologies.}
	\label{fig:controls_moped}
\end{figure}

Finally, Fig.~\ref{fig:controls_scooter} shows the e-scooter to have a very small energy recuperation potential, since the ratio of circulative-to-dissipative energy is significantly lower than for conventional cars~\cite{GuzzellaSciarretta2007} and the kinetic energy of the e-scooter is mostly dissipated via drag forces, resulting in a very small negative motor power and increase in $E_\mathrm{b}$. Due to the low energy recuperation potential, one could even consider to reduce the TCO by removing the power electronics which enable energy regeneration and hence reduce the vehicle's weight.
Following a similar trend, Fig.~\ref{fig:controls_moped} shows that the influence of recuperating with an e-moped is also minimal as the increase in $E_\mathrm{b}$ during the braking phase is negligible. However, the higher mass and velocity result in a larger kinetic energy and higher negative motor power. Therefore, the potential energy savings might be able to compensate for the cost of a regenerative braking system for driver masses above a certain value. Following this idea, further research could include a sensitivity analysis on the influence of energy recuperation on the TCO with respect to the total weight of the electric moped.




\subsection{Validation}
We validate the accuracy of our convex models by simulating our optimal results with a nonlinear vehicle model based on the original EM and battery data.
Thereby, the difference in consumption between both models for all vehicles and transmission technologies lies within a range of 0.42--1.00\%, validating the accuracy of our optimization framework.


	\section{Conclusion}\label{sec:conclusion}

In this paper, we explored an efficient approach to optimize the design of electric micromobility vehicles with respect to their total cost of ownership (TCO). 
In particular, we focused on e-scooters equipped with a fixed-gear transmission (FGT) and e-mopeds equipped with an FGT or a continuously variable transmission (CVT).
To this end, we derived a partially-convex model of their powertrains, defined application-specific performance requirements, generated custom driving cycles, and formulated an optimization problem in which the components' size and the control of the powertrain are jointly optimized. 
In order to achieve convergence in a reasonable time, we devised an iterative algorithm based on second-order conic programming that exploits the partial convexity of the problem.

Our numerical case studies investigated the impact of the vehicles' employment area on the optimal solution, revealing that the presence of hills would result in an increase of TCO and components' cost of almost 10\%, with respect to flat regions. This clearly underlines the importance of accounting for the application terrain when designing and deploying micromobility vehicles.
What is more, our results indicated that, given the relatively low weight of such vehicles---and in line with the state of the art---enabling regenerative braking capabilities would not significantly improve the achievable performance, whilst using a CVT may even increase the overall TCO.


This work opens the field for the following extensions:
First, we would like to extend our framework to accommodate hybrid-human vehicles such as e-bikes.
Second, we would like to embed our approach within the system-level design of Mobility-as-a-Service systems such as intermodal Autonomous Mobility-on-Demand (AMoD), as they would clearly benefit from the presence of micromobility vehicles as a mode of transportation~\cite{Wollenstein-BetechSalazarEtAl2021}.
	
	\section*{Acknowledgment}
	\noindent
	The authors thank Dr. Ilse New for proofreading this paper and Juriaan van den Hurk for the fruitful discussions.
	
	\ifextendedversion
	\appendix
\subsection{Simulation Parameters}
\label{Sec:simpar}
The values of the constants used for the simulation of both vehicles are given in Table~\ref{Tab:Simulation Parameters} and are based on the values from~\cite{EScooterGuideTest2021,VerbruggenSalazarEtAl2019,Ulrich2005,ruan2018development} and~\cite{CiezWhitacre2017, KochhanFuchsEtAl2017, Pricewise2021,Blum2021,Childers2000,SuperSocoCux2021,SuperSoco2021}.
\begin{small}
	\begin{table}
		\centering
		\caption{Simulation parameters.}
		\label{Tab:Simulation Parameters}\scriptsize
		\begin{tabular}{l l l l } \hline 
			\textbf{Parameter [unit]}  & \textbf{Symbol} & \textbf{Scooter} & \textbf{Moped}   \\ \hline
			Driver mass [\si{kg}]&  $m_\mathrm{d}$  &  75  &  75      \\ 
			Rolling resistance coefficient [-]         &  $c_\mathrm{rr}$  &  0.03  &  0.015   \\ 
			Gravitational constant [\si{m/s\textsuperscript{2}}]       &  $g$  &  9.81  &  9.81    \\ 
			Air density [\si{kg/m\textsuperscript{3}}]         &  $\rho_\mathrm{a}$  &  1.225  &  1.225   \\ 
			Frontal area [\si{m\textsuperscript{2}}]        &  $A_\mathrm{f}$  &  0.68  &  0.7 \\ 	
			Air drag coefficient [-]        &  $ c_\mathrm{d}$  &  1  &  0.7     \\ 
			CVT ratio coverage [-]        &  $c_\mathrm{f}$  &  -  &  2.7     \\ 
			Final drive efficiency (FGT) [-]         &  $\eta_\mathrm{fd}$  &  1  &  1     \\
			Final drive efficiency (CVT) [-]         &  $\eta_\mathrm{fd}$  &  -  &  0.97     \\
			Gearbox efficiency (FGT) [-]    &  $\eta_\mathrm{gb}$  &  0.97  &  0.97     \\	
			Gearbox efficiency (CVT) [-]     &  $\eta_\mathrm{gb}$  &  -  &  0.88     \\
			Final drive ratio [-]    &  $\gamma_\mathrm{fd}$  &  1  &  1    \\
			Brake fraction [-]    &  $R_\mathrm{b}$  &  0.5  &  0.5     \\ 
			Effective rolling radius [\si{m}]   &  $r_\mathrm{w}$  &  0.125  &  0.193    \\ 
			Friction coefficient [-]   &  $\mu_\mathrm{x}$  & 0.4   & 0.4     \\	
			Maximum EM speed [\si{rad/s}]    &  $\omega_\mathrm{em,max}$  &  600  &  600     \\
			Auxiliary power [\si{W}]    &  $P_\mathrm{aux}$  &  10  &  10    \\ 
			Minimum state-of-charge [-]    &  $\zeta_\mathrm{b,min}$  &  0.2  &  0.2   \\ 
			Maximum state-of-charge  [-]   &  $\zeta_\mathrm{b,max}$  &  1  &  1   \\
			Frame mass  [\si{kg}]   &  $m_\mathrm{f}$  &  10  &  60   \\
			Base mass CVT [\si{kg}]    &  $m_\mathrm{cvt,base}$  & -  & 0.5     \\ 
			Range until battery end-of-life [\si{km}]    &  $D_\mathrm{max}$  & 8000  & 120000   \\ 
			\textit{Specific masses:}     &   &   &    \\
			EM [\si{kg/kW}]   &  $\rho_\mathrm{em}$  &  0.5  & 0.5   \\
			Battery [\si{kg/kWh}]  &  $\rho_\mathrm{bat}$  &  4.73  & 4.73     \\ 
			FGT [\si{kg}]   &  $\rho_\mathrm{FGT}$  & 0.01   & 0.075     \\
			CVT [\si{kg}]    &  $\rho_\mathrm{cvt}$  & -  & 0.05    \\
			\textit{Specific costs:}     &   &   &    \\
			Electricity [\si{\euro/kWh}]     &  $c_\mathrm{el}$  & 0.22  & 0.22   \\
			Battery [\si{\euro/kWh}]   &  $c_\mathrm{bat}$  & 285  & 285    \\
			EM (FGT) [\si{\euro/kW}]   &  $c_\mathrm{em}$  & 101  & 101    \\
			EM (CVT) [\si{\euro/kW}]   &  $c_\mathrm{em}$  & -  & 150    \\
			Additional costs [\si{\euro}]  &  $c_\mathrm{add}$  & 88  & 209     \\
			\textit{Performance requirements:}     &   &   &    \\
			Time to maximum speed [\si{s}]   &  $t_\mathrm{acc}$  & 7.5   & 11     \\
			Start gradient [\%]   &  $\theta_\mathrm{start}$  & 10   & 20    \\ 
			Expected range [\si{km}]   &  $D_\mathrm{exp}$  & 25   & 100     \\
			Maximum speed [\si{km/h}]   &  $v_\mathrm{max}$  & 25   & 45    \\ \hline
		\end{tabular}
	\end{table}
\end{small}

\subsection{Convergence Analysis on Vehicle Mass}
As mentioned in Section~\ref{Sec:Discussion}, we have no global optimality guarantees for the solution found by Algorithm~1. However, we can perform a numerical analysis to determine whether Algorithm~1 approaches a unique solution. This is done for a fixed EM size, different initial masses $m_\mathrm{v,0}$ and a tolerance $\epsilon$ of \unit[10$^{-6}$]{kg} for both vehicles and scenarios. The results presented in Fig.~\ref{fig:convergence_analysis_sooter} show that the vehicle mass of the scooter on the flat road converges to a unique value of \unit[12.7]{kg} within a tolerance of 0.05\%. Moreover, we obtain similar results for the e-scooter in the hills  and e-moped with the FGT and CVT transmission in both scenarios.

\begin{figure}[!t]
	\centering
	\includegraphics[ width=1\columnwidth]{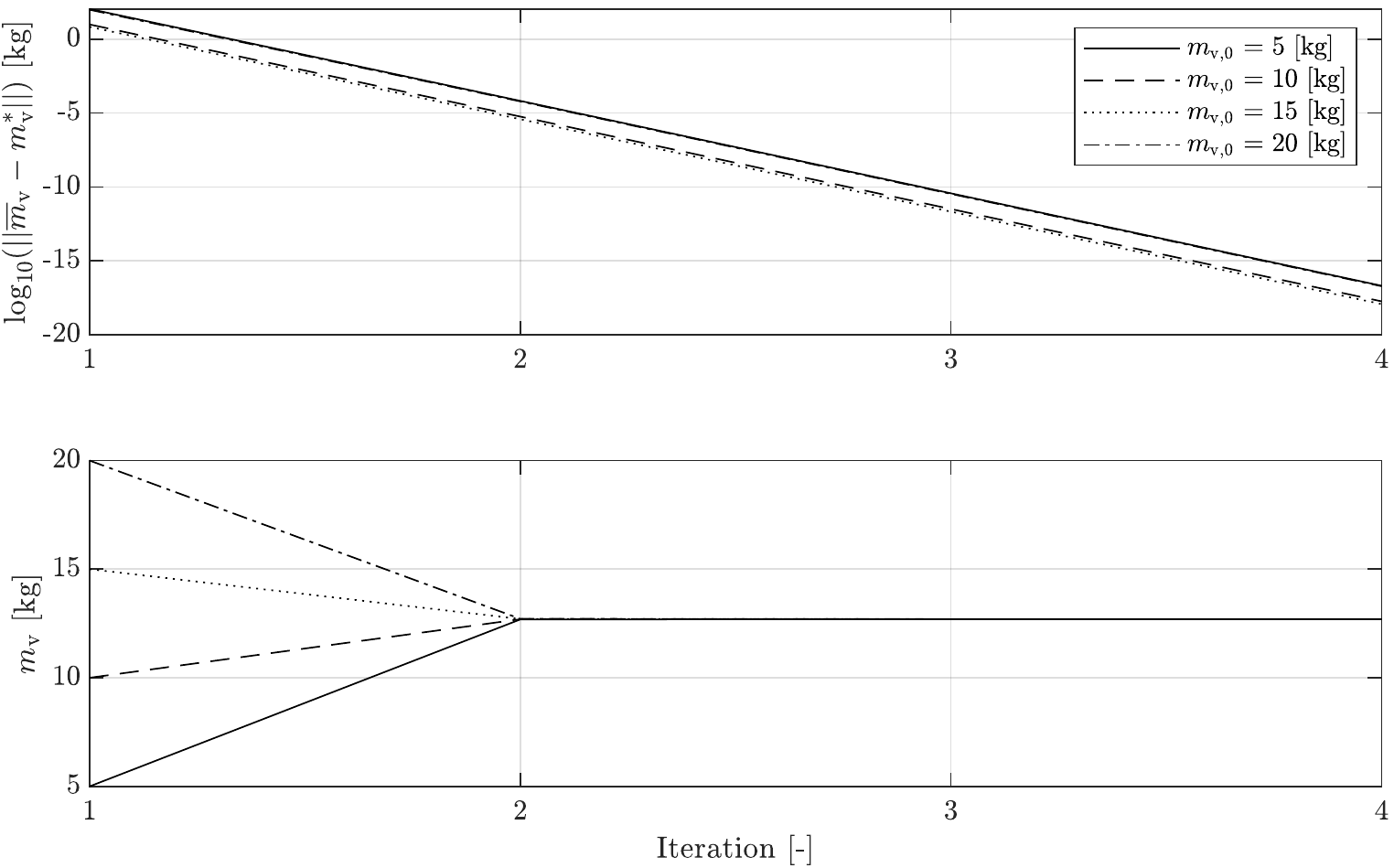}
	\caption{Convergence analysis of the vehicle mass of the e-scooter on the flat road, using a \unit[590]{W} EM.}
	\label{fig:convergence_analysis_sooter}
\end{figure}

\subsection{Comparison of Transmission Technologies}
This section elaborates on the difference between the FGT and CVT transmissions in the e-moped. Apart from the differences observed in the gear ratio, we also notice that the FGT vehicle's operating points lie in more low-efficiency regions compared to those of the CVT vehicle, as shown in Fig. \ref{fig:op_points_moped_hills}.
This difference is caused by the performance requirements and maximum torque limits, since the FGT can select only one gear ratio to comply to the constraints, whereas the CVT has a variable ratio to meet all constraints, which allows the EM to operate at higher speeds in a region that is more efficient. This effect is shown in more detail in Fig.~\ref{fig:controls_moped_w_gamma}, where the gear ratio and EM speed of the CVT are higher than those of the FGT for most time instances.
We also note that a share of the CVT operating points is not located in the most efficient regions, which is caused by the EM speed constraint in \eqref{eq:speed} and limited CVT ratio coverage.
\begin{figure}[!t]
	\centering
	\includegraphics[ width=1\columnwidth]{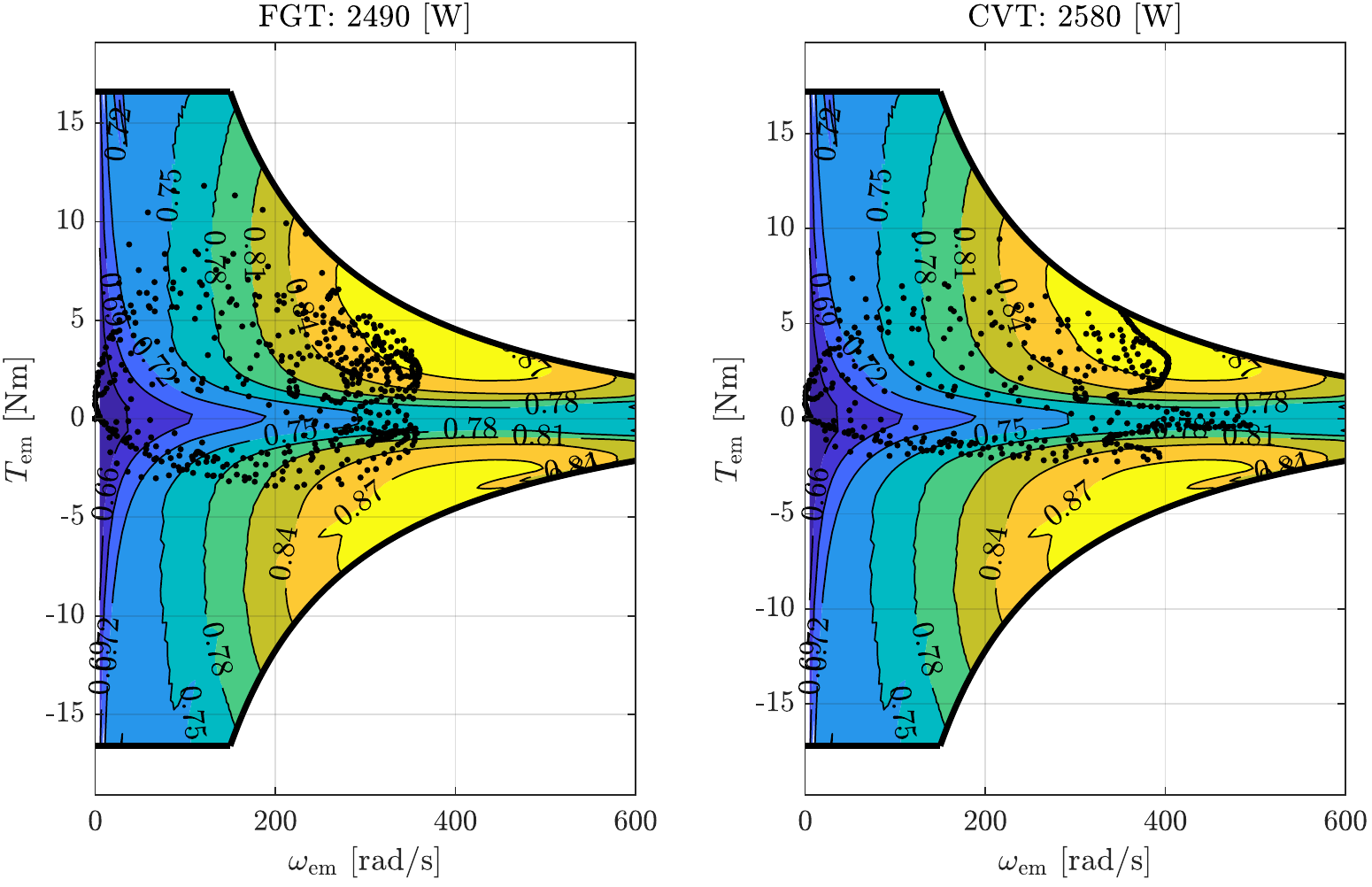}
	\caption{Operating points (black dots) in convex efficiency map of the FGT (left) and CVT (right) moped in the hills.}
	\label{fig:op_points_moped_hills}
\end{figure}
\begin{figure}[!t]
	\centering
	\includegraphics[ width=1\columnwidth]{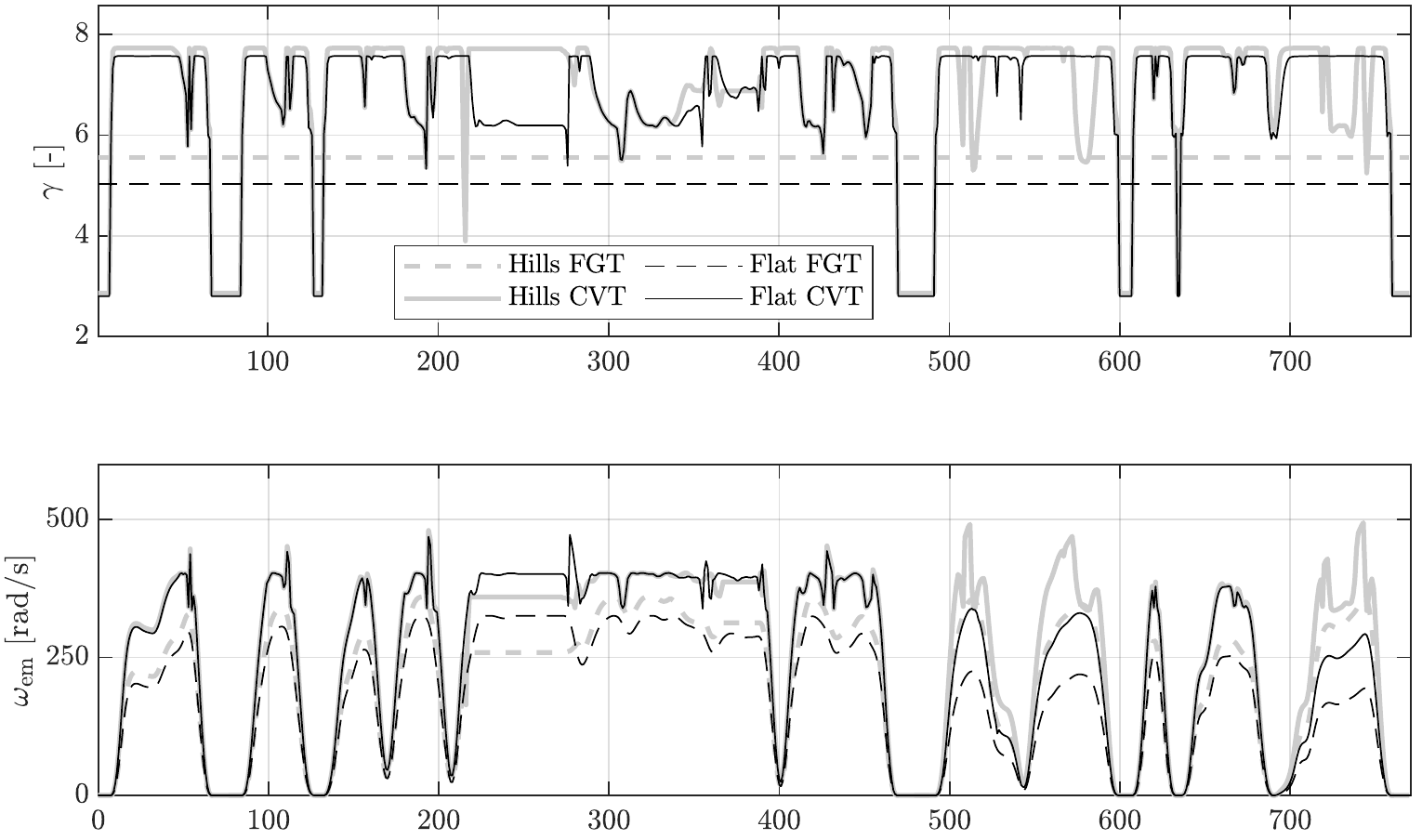}
	\caption{Gear ratio and EM speed of the e-moped for both scenarios and transmission technologies.}
	\label{fig:controls_moped_w_gamma}
\end{figure}

	\fi
	

	
	
	%
	
	\bibliographystyle{IEEEtran}        
	\bibliography{../../../Bibliography/main,../../../Bibliography/SML_papers}

\end{document}